 \documentclass[final,5p,times,twocolumn]{elsarticle}

\usepackage{graphicx}
\usepackage{amssymb}
\usepackage{amsthm}
\usepackage{amsmath}
\usepackage{color}

\usepackage{bm}
\usepackage{lineno}

\usepackage{threeparttable}

\journal{Journal Name}

\begin{document}

\begin{frontmatter}


\title{Transverse asymmetry of $\gamma$ rays from neutron-induced compound states of ${}^{140}{\rm La}$}



\author[1]{T. Yamamoto}
\author[2]{T. Okudaira}
\author[1,2]{S. Endo}
\author[3]{H. Fujioka}
\author[1]{K. Hirota}
\author[4]{T. Ino}
\author[1]{K. Ishizaki}
\author[2]{A. Kimura}
\author[1]{M. Kitaguchi}
\author[5]{J. Koga}
\author[5]{S. Makise}
\author[1]{Y. Niinomi}
\author[2]{T. Oku}
\author[2]{K. Sakai}
\author[6]{T. Shima}
\author[1]{H. M. Shimizu}
\author[5]{S. Takada}
\author[3]{Y. Tani}
\author[6]{H. Yoshikawa}
\author[5]{T. Yoshioka}

\address[1]{Nagoya University, Furocho, Chikusa, Nagoya 464-8602, Japan}
\address[2]{Japan Atomic Energy Agency, 2-1 Shirane, Tokai 319-1195, Japan}
\address[3]{Tokyo Institute of Technology, Meguro, Tokyo 152-8551, Japan}
\address[4]{High Energy Accelerator Research Organization KEK 1-1 Oho, Tsukuba, Ibaraki 305-0801, Japan}
\address[5]{Kyushu University, 744 Motooka, Nishi, Fukuoka 819-0395, Japan}

\address[6]{Research Center for Nuclear Physics, Osaka University, Ibaraki, Osaka 567-0047, Japan}

\begin{abstract}
A correlation term ${{\bm \sigma}_{n} }\cdot ({\bm k_{n}\times \bm k_\gamma}) $ in the ${}^{139}{\rm La}(\vec{n},\gamma)$ reaction has been studied utilizing epithermal polarized neutrons and germanium detectors. 
The transverse asymmetry for single $\gamma$-ray transition was measured to be $0.60\pm0.19$ in the $p$-wave resonance.

\end{abstract}

\begin{keyword}
Nuclear physics
\end{keyword}

\end{frontmatter}


\section{INTRODUCTION}
Since the discovery of large parity violation (P violation) in the neutron absorption reaction of $^{139}{\rm{La}}$ in the 0.74~eV $p$-wave resonance by Dubna's group \cite{Alfimenkov}, many systematic studies have been performed on P violation in compound nuclear reactions \cite{Mitchell}. These phenomena have been interpreted as 
the result of the interference between partial waves with opposite parities ($sp$-mixing model) \cite{Flambaum}. In addition, it is suggested that this enhancement is applicable not only to P violation but also to time reversal symmetry violation (T violation) \cite{Gudkov}. Detailed studies of the enhancement mechanism on the basis of the $sp$-mixing model are necessary to evaluate the feasibility of highly sensitive T violation search experiments \cite{Bowman}.\\ 
 As a first step, Okudaira et al. measured the energy-dependent angular distribution of $\gamma$ rays consistently with the $sp$-mixing model in the vicinity of the 0.74~eV $p$-wave resonance in $^{139}{\rm La}(n,\gamma){}^{140}{\rm La}$ reaction \cite{Okudaira}. Further studies are necessary to uniquely determine the contributions of individual partial amplitudes in the entrance channel.
In this paper, we report the first result of the measurement of the energy dependence of the $\gamma$-ray asymmetry with respect to the transverse polarization of incident neutrons.
 
\section{EXPERIMENT}
\subsection{Experimental setup}
\subsubsection{BL04 ANNRI}
\begin{figure}[h]
  \centering
  \includegraphics[width=7cm]{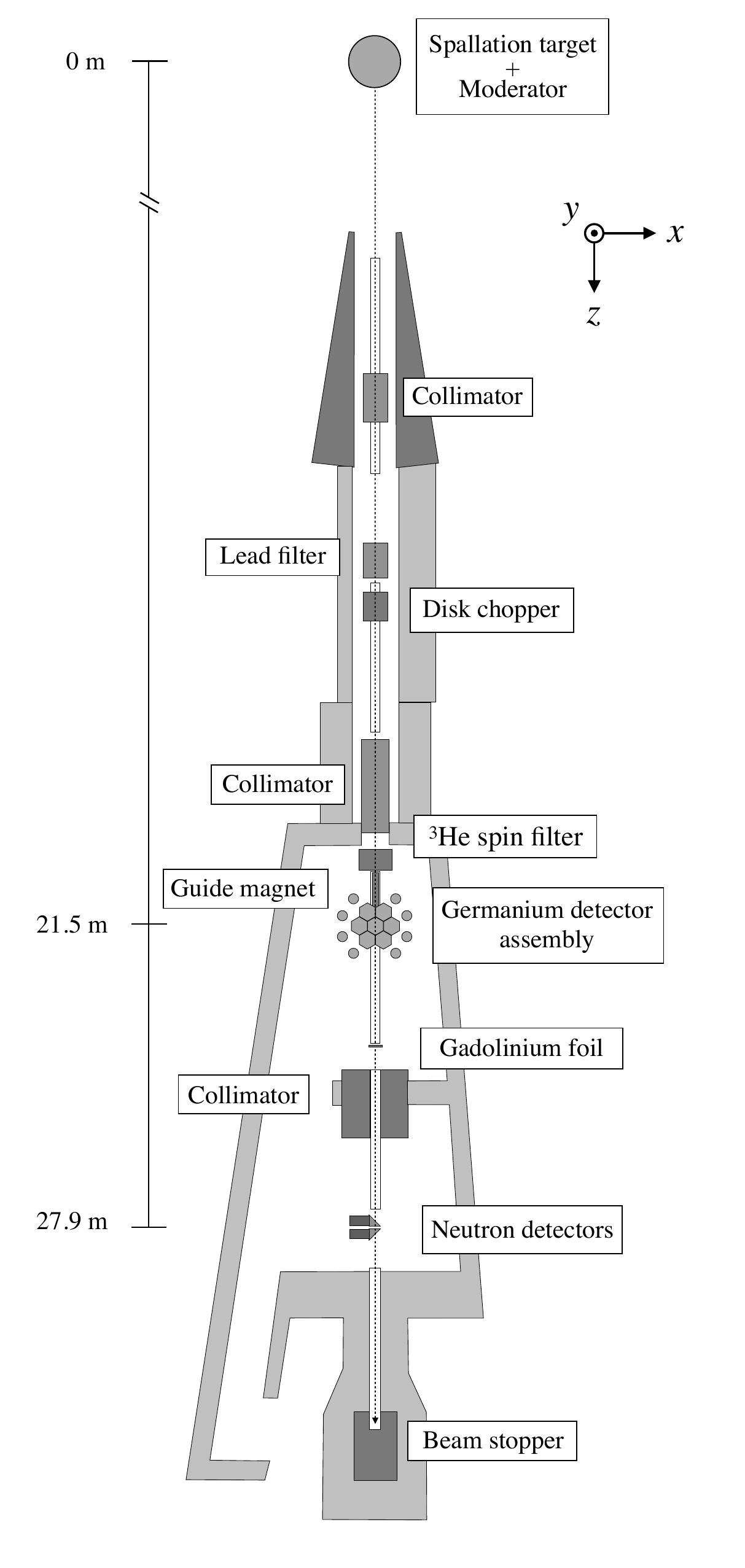}
  \caption{Top view of BL04 ANNRI.}
  \label{fig:beamline}
\end{figure}

This measurement was performed at the Accurate Neutron-Nucleus Reaction Measurement Instrument (ANNRI) installed at beam line 04 (BL04) of the Material and Life Science Experiment Facility (MLF) of the Japan Proton Accelerator Research Complex (J-PARC). In MLF, pulsed neutrons are generated by the spallation reaction induced by 3~GeV proton beam supplied by the Rapid-Cycling Synchrotron. The repetition rate was 25~Hz and an average beam power was 500~kW during the measurement. Neutrons produced were then moderated by a liquid hydrogen moderator and supplied to the beam line.
As shown in Fig.\ref{fig:beamline}, the neutron beam was collimated to a 15-mm-diameter spot at the target position, placed 21.5~m from the moderator surface \cite{Kino}.
The lead filter (37.5mm-thick in total) was installed to suppress the background caused by fast neutrons and $\gamma$ flashes.
The disk choppers operated in synchronization with the timing of the proton beam incidence to avoid frame overlap due to low-energy neutrons.
A ${\rm{^{3} He}}$ spin filter as a neutron polarizer was installed 1.5~m upstream from the target.
The germanium detector assembly was arranged surrounding the target to detect the $\gamma$ rays from the $(n, \gamma)$ reaction.
Neutrons passing through the target were detected by neutron detector located 27.9~m from the moderator through a downstream collimator. To avoid saturation of the neutron detector, a 10~$\mu$m-thick gadolinium foil was placed 3~m upstream from the neutron detector. 

\subsubsection{$\gamma$-ray detectors}
\begin{figure}[h]
\hspace*{-1.5cm}
  \centering
  \includegraphics[width=11cm]{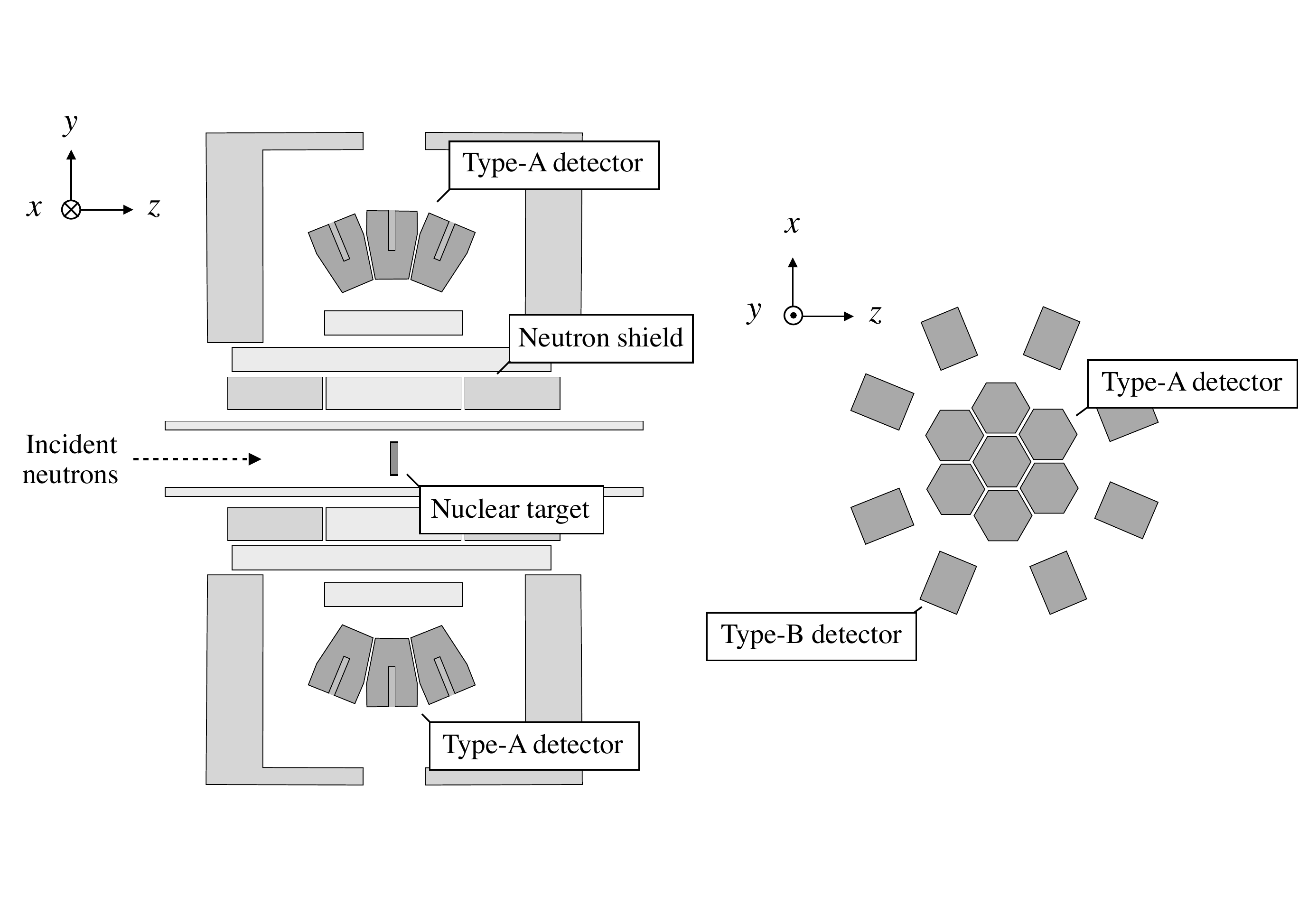}
  \caption{Germanium detector assembly. The left figure is a sectional view and the right figure is a top view.}
  \label{fig:Gedet}
\end{figure}

\begin{figure}[h]
  \centering
  \includegraphics[width=9cm]{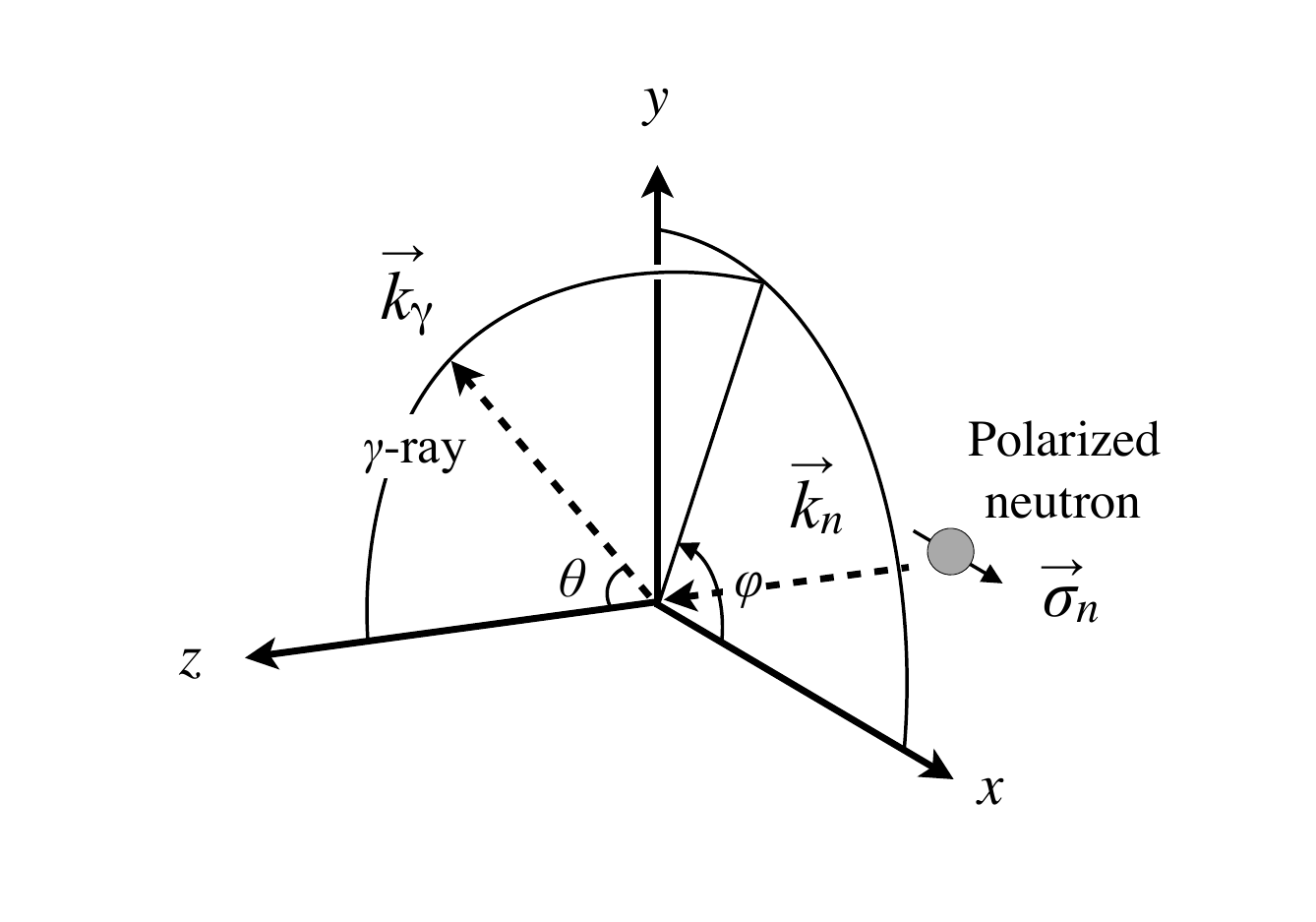}
  \caption{Definition of coordinate systems and vector elements. The $z$ axis is defined in the beam direction, the $y$ axis is the vertical upward axis,  the $x$ axis is perpendicular to them.
  The polar angle in the spherical coordinate is denoted by $\theta$ and the azimuthal angle is $\varphi$. }
  \label{fig:Defsys}
\end{figure}

Germanium detector assembly shown in Fig.\ref{fig:Gedet} was used to detect the $\gamma$ rays emitted from the target \cite{Kimura}.
A right handed system is used in this paper as shown in Fig.\ref{fig:Defsys}, and the neutron beam direction is defined as the $z$ axis. 

The germanium detector assembly consists of two kinds of detectors. One is called a ``A-type detector'', which are located above and below the target, respectively. Each A-type detector has seven separated germanium crystal region. These crystals are placed at $\theta = 71^{\circ}, 90^{\circ},$ and $109^{\circ}$.
The other is called a ``B-type detector'', which consists of eight germanium crystals surrounding the target. These crystals are located at $\theta = 36^{\circ}, 72^{\circ}, 108^{\circ}$, and $144^{\circ}$ \cite{Takada}.
 During the experiment, germanium crystals were cooled to 77~K by liquid nitrogen or an mechanical refrigerator. The energy threshold level for $\gamma$-ray detection was set at approximately 100~keV in order to avoid electronic noise. In this analysis, we used only the lower A-type detectors because the upper A-type detectors' condition was unstable. 

\subsubsection{Neutron detectors}
Two types of lithium glass scintillation detectors were used to detect neutrons transmitted through the target \cite{Nakao}. One was a ${\rm{^{6}Li}}$ 90$\%$ enriched type (GS20) and the other was a ${\rm{^{7}Li}}$ 99.99$\%$ enriched type (GS30). The size of both scintillators is $\rm 50~mm \times50~mm \times 1~mm$. The ${\rm{^{6}Li}}$ isotope has a neutron absorption cross section of 940~barn for 25~meV neutron. On the other hand ${\rm{^{7}Li}}$ has a very low sensitivity (0.045~barn for 25~meV neutron) for neutron detection. Since GS20 and GS30 have similar sensitivities for $\gamma$ rays, the $\gamma$-ray background of GS20 was evaluated by GS30 by installing two detectors in close proximity.

\subsubsection{Neutron polarizer}
\begin{figure*}[h]
  \centering
  \includegraphics[width=16cm]{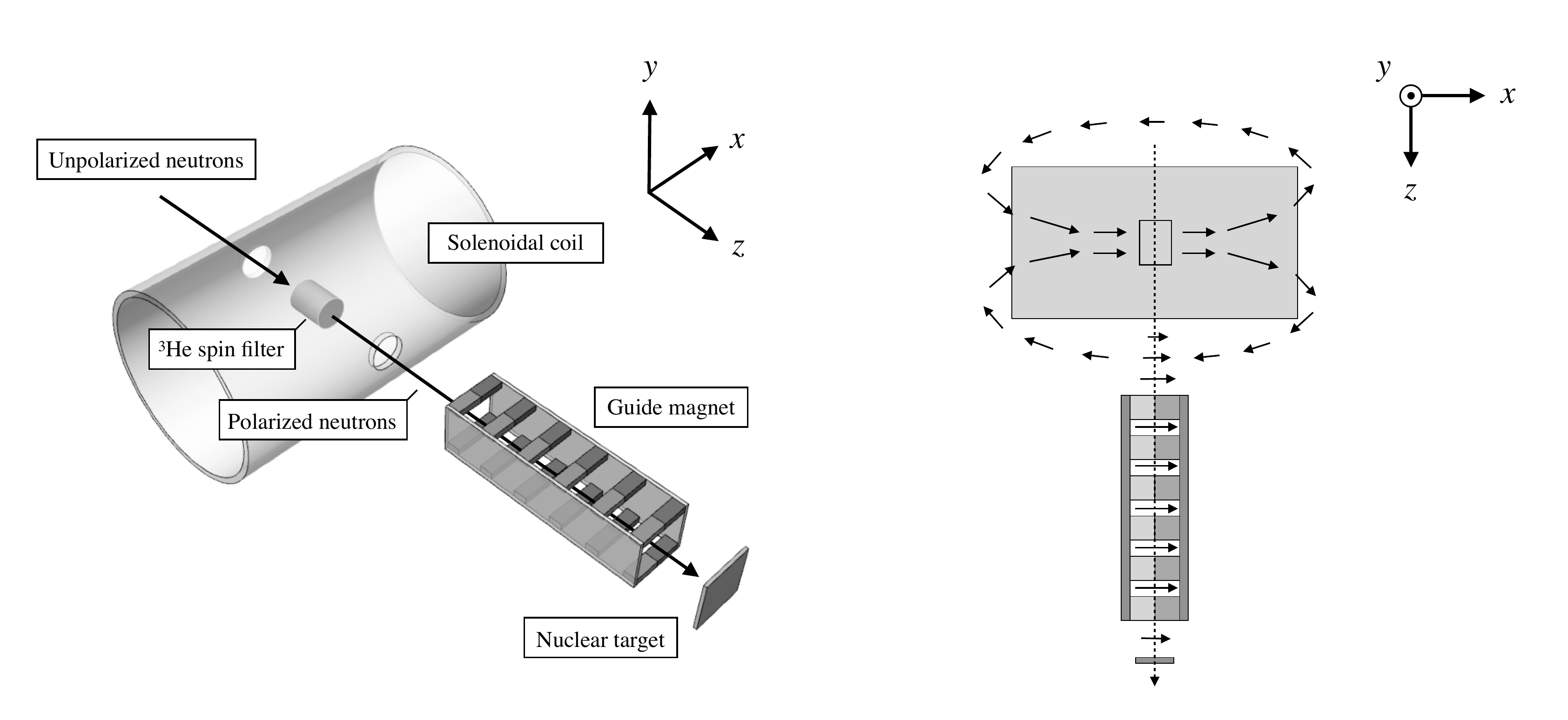}
  \caption{${\rm{^{3} He}}$ spin filter and a guide magnet for neutron spin transport. The left figure is a solid view of the system. Unpolarized neutrons were injected from side of the solenoid coil, and they were polarized by passing through the ${\rm {^{3}He}}$ spin filter installed in the solenoid coil, and they were irradiated to the downstream nuclear target. The solenoid coil is covered with a permalloy film to prevent the influence of the external magnetic field. The right figure is a top view of the system. Solid arrows indicate the direction of the magnetic field and a dashed arrow indicates the direction of the neutron beam. The magnetic field antiparallel to $x$ axis  outside the solenoid coil was canceled by the guide magnet composed of permanent magnets, and in addition, the magnetic field was applied in the $x$ axis parallel direction. Therefore, polarized neutrons always maintained polarization due to adiabatic transport in the guiding magnetic fields. }
  \label{fig:Magcond}
\end{figure*}

A ${\rm{^{3}He}}$ spin filter was used as the neutron polarizer.
The ${\rm{^{3}He}}$ spin filter is a glass cell of 50-mm-diameter and 77-mm-length filled with ${\rm{^{3}He}}$ gas to about 3~atm. Neutrons were polarized by passing through the polarized ${\rm{^{3}He}}$ which has a large spin-dependent neutron cross section: the absorption cross sections for 25~meV neutrons with spins parallel and antiparallel to ${\rm{^{3}He}}$ are approximately 0 and 10666~barn, respectively. The  ${\rm{^{3}He}}$ was polarized by the spin exchange optical pumping (SEOP) method \cite{SEOP}. A solenoid coil was used to define the quantization axis and to suppress the relaxation of ${\rm {^{3}He}}$ polarization due to the nonuniformity of the external magnetic field. In this experiment, up and down polarized neutrons are defined as parallel and antiparallel to the $x$ axis, respectively.
A magnetic field of approximately 15~G was applied by the solenoid coil at the cell installation position. In order to polarize neutrons along the $x$ axis, after ${\rm {^{3}He}}$ was polarized by SEOP, the glass cell was rotated adiabatically 90~degrees in the solenoid coil.
Since the spin of ${\rm {^{3}He}}$ follows the magnetic field of the solenoid coil, ${\rm {^{3}He}}$ is polarized in the $x$ axis direction. The spin filter was installed into the beam line after the ${\rm {^{3}He}}$ was polarized in the ${\rm {^{3}He}}$ polarization station in MLF outside the beam line \cite{NSF}. As shown in Fig.\ref{fig:Magcond}, unpolarized neutrons were injected from side of the solenoid coil. After passing through the spin filter, a guide magnetic field was applied from the ${\rm {^{3}He}}$ spin filter to the target using a permanent magnet to keep the polarization of neutrons.

\begin{figure}[h]
  \centering
  \includegraphics[width=9cm]{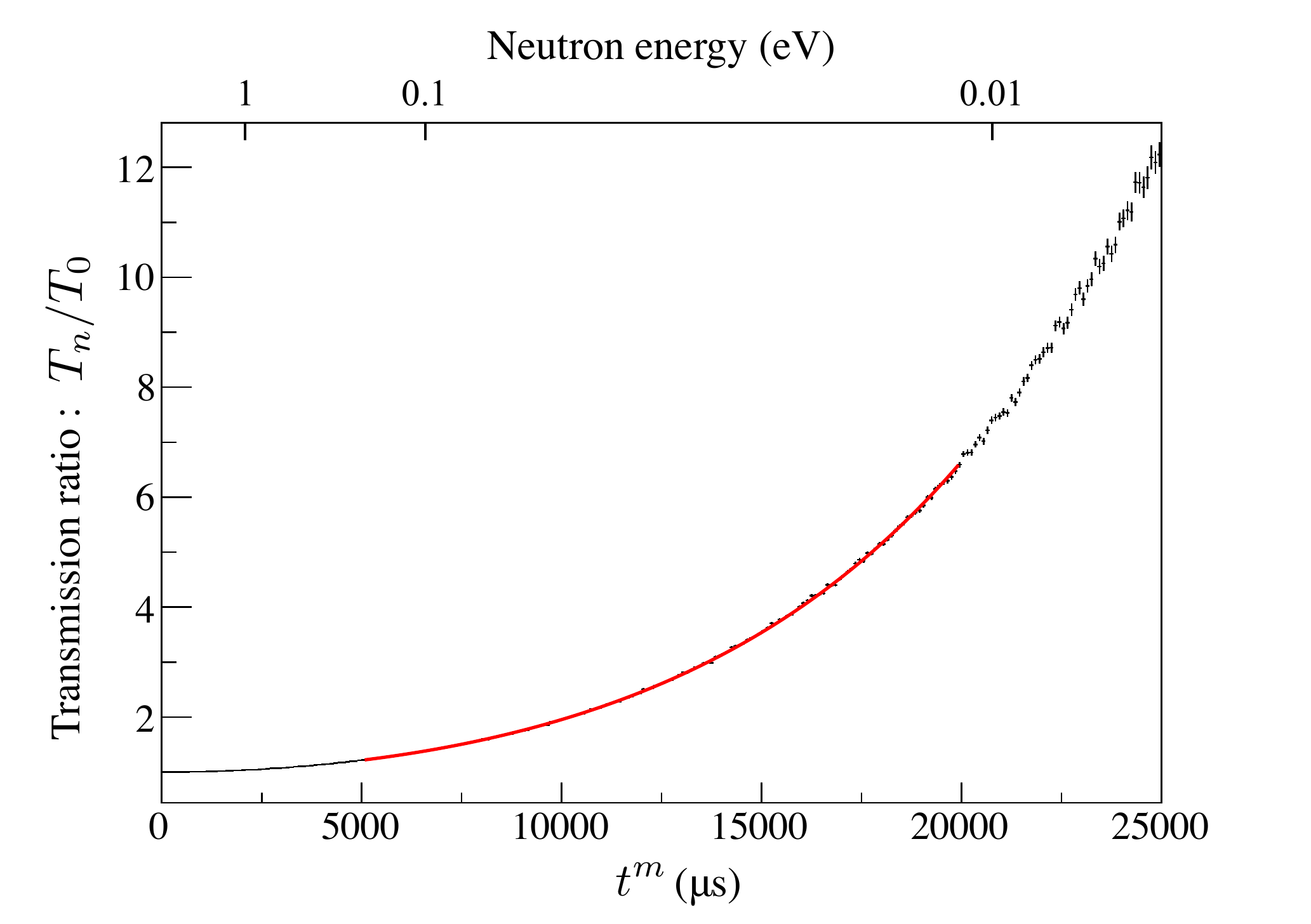}
  \caption{Ratio of neutron transmission through polarized ${\rm {^{3}He}}$ to that through unpolarized ${\rm {^{3}He}}$. The solid line is the result of fitting by $T_{n}/T_{0} = \cosh(P_{\rm {He}}n_{\rm {He}}t\Delta\sigma)$. From this result, the neutron polarization can be determined.}
  \label{transmission}
\end{figure}

The neutron transmittance of the ${\rm{^{3}He}}$ spin filter $T_{n}$ depends on the ${\rm {^{3}He}}$ polarization ratio $P_{\rm {He}}$ and is written as
\begin{equation}
\label{transpol}
T_{n} = T_{0}\cosh(P_{\rm {He}}n_{\rm {He}}t\Delta\sigma),
\end{equation}
where $T_{0}$ is the neutron transmittance for unpolarized ${\rm{^{3}He}}$, $n_{\rm {He}}$ is the atomic number density of ${\rm{^{3}He}}$, $t$ is the cell length, $\Delta\sigma$ is the neutron absorption cross section of ${\rm{^{3}He}}$. The value of $\Delta\sigma$ was extrapolated from the cross sections at 25~meV, assuming the energy dependence of the $1/v$ law. The value of $n_{\rm He}t$ was evaluated as $n_{\rm He}t = 19.331 \pm 0.026$~atm$\cdot$cm from the ratio of transmittance of the vacuum glass cell and the unpolarized ${\rm{^{3} He}}$. As shown in Fig.\ref{transmission}, the polarization ratio of ${\rm {^{3} He}}$ was evaluated by fitting using Eq.\ref{transpol}. 

\begin{figure}[h]
  \centering
  \includegraphics[width=9cm]{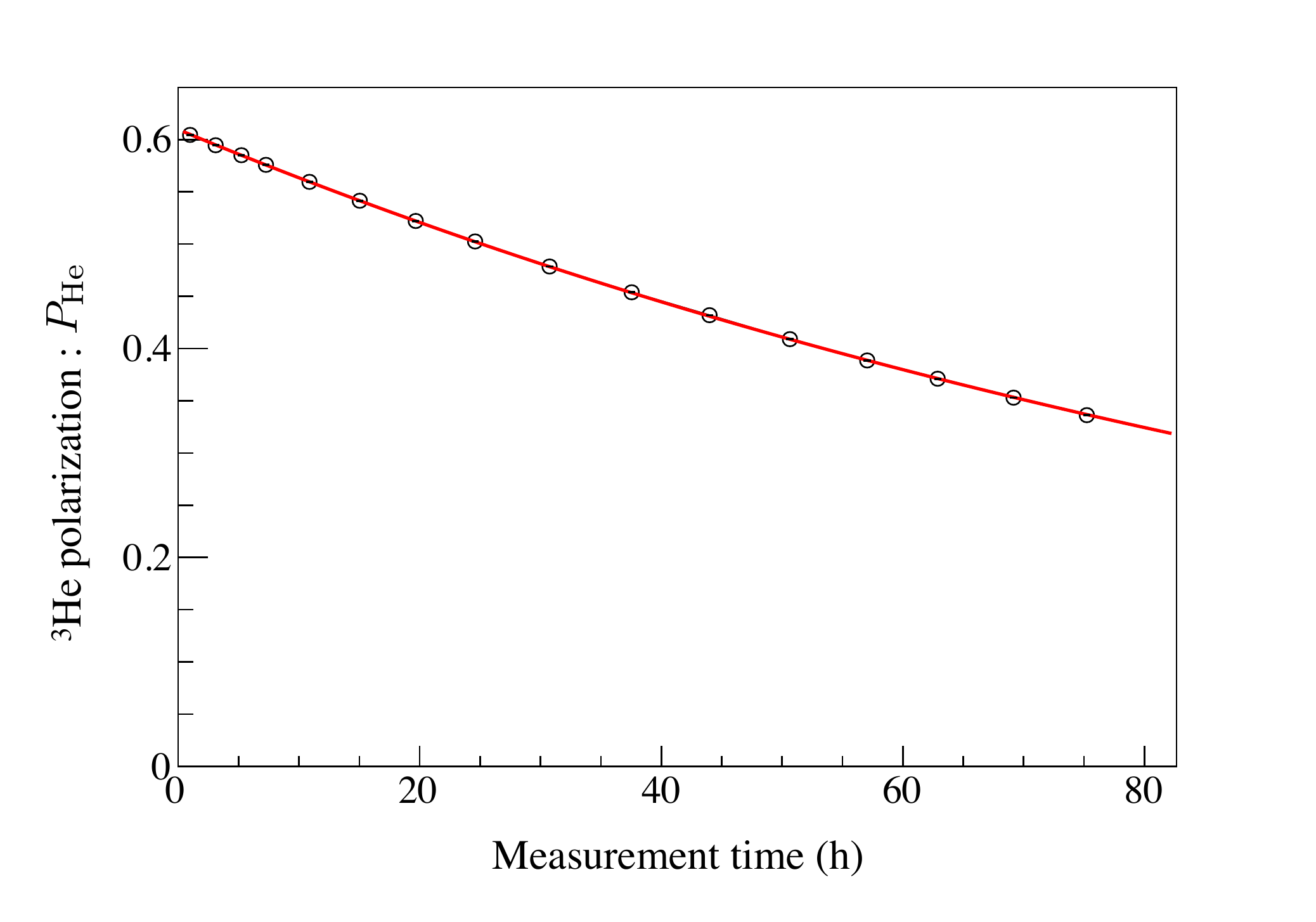}
  \caption{Time dependence of the polarization ratio of ${\rm {^{3}He}}$. The polarization of ${\rm {^{3}He}}$ relaxed with time. The solid line is the fitting result of the relaxation time constant using the exponential decay function.}
  \label{ref:decay}
\end{figure}

Figure \ref{ref:decay} shows the time dependence of the polarization ratio of ${\rm {^{3} He}}$ from the installation of the ${\rm {^{3}He}}$ spin filter onto the beamline to the end of the measurement. The polarization ratio of ${\rm {^{3} He}}$ at the start of the measurement was estimated to be approximately 60$\%$, and the relaxation time of ${\rm {^{3} He}}$ was estimated to be 127.0 $\pm$ 0.1~hours. The neutron polarization ratio $P_{n}$ and the ${\rm{^{3}He}}$ polarization ratio $P_{\rm{He}}$ are related through
\begin{equation}
\label{polpol}
P_{n}=\tanh(P_{\rm{He}}n_{\rm {He}}t\Delta\sigma).
\end{equation}
The neutron polarization ratio was derived using ${\rm{^{3}He}}$ polarization.
This value was used to correct the asymmetry analysis in the following section.

\subsubsection{Data acquisition system}
Since germanium detectors have a high energy resolution, they also require a high analog to digital converter (ADC) resolution. A V1724, 8~channels, 14~bit, 100~MS/s, peak hold digitizer manufactured by CAEN was used as the ADC for the germanium detectors \cite{CAEN}.
The neutron detector require a fast response to the ADC because a high count rate was assumed for high flux neutron beam. A V1720, 8~channels, 12~bit, 250~MS/s, charge sensitive digitizer was used as the ADC for neutron detectors.
Both ADCs distribute the input signal to the pulse height processing system and the timing processing system. 
In the timing processing system, the difference between proton incident timing and the detection time is recorded as time information $t^{m}$.
In this measurement, we can calculate the neutron energy $E_n$ applying the time-of-flight (TOF) method and the pulse height information obtained in V1724 correspond to $\gamma$-ray energy $E_{\gamma}$ after energy calibration.

\subsection{Measurement}
The target was a metal lanthanum plate of  $\rm 40~mm \times40~mm \times 3~mm$ with a purity of 99.9$\%$. Measurements were performed in March and May in 2019. The total measurement time was 77.7 and 76.5 hours in the two opposite spin states.
The spin of ${\rm{^{3}He}}$ was flipped approximately every 4~hours using the adiabatic fast-passage (AFP) NMR method. After the polarization measurement, the ${\rm {^{3}He}}$ was depolarized and the neutron transmittance was measured to evaluate the polarization ratio of ${\rm {^{3}He}}$. Neutron transmittance of the evacuated glass cell was also measured to isolate the ${\rm {^{3}He}}$ gas contribution during ${\rm {^{3}He}}$ spin filter usage. A melamine ($\rm{C_{3}H_{6}N_{6}}$) target was used to correct the $\gamma$-ray detection efficiency of each germanium detector using $\gamma$ rays emitted from the $^{14}{\rm {N}}(n,\gamma)$ reaction. 
\section{DATA ANALYSIS AND RESULTS}

\subsection{Transverse asymmetry}
\begin{figure}[h]
  \centering
  \includegraphics[width=9cm]{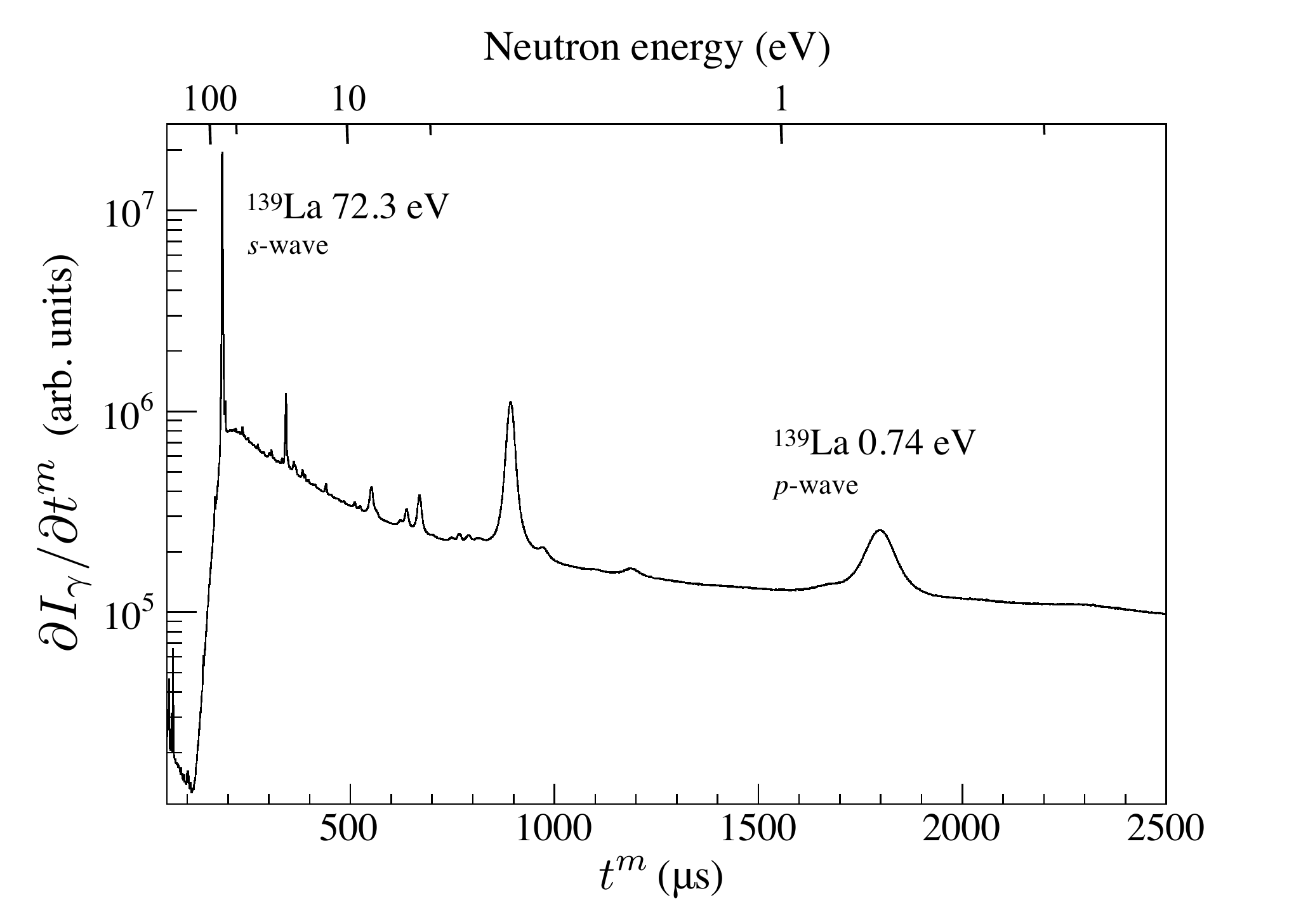}
  \caption{Neutron TOF spectrum of $\gamma$ rays from the $(n,\gamma)$ reaction with the lanthanum target. The resonance at around $t^{m} \sim$ 1800~$\mu$s and 200~$\mu$s are the $p$-wave resonance of $^{139}{\rm {La}}$ and the $s$-wave resonance of $^{139}{\rm {La}}$, respectively. The small bump around $t^{m} \sim$ 1700~$\mu$s derives from the resonance of $^{149}{\rm {Sm}}$.}
  \label{tof}
\end{figure}

\begin{figure}[h]
  \centering
  \includegraphics[width=9cm]{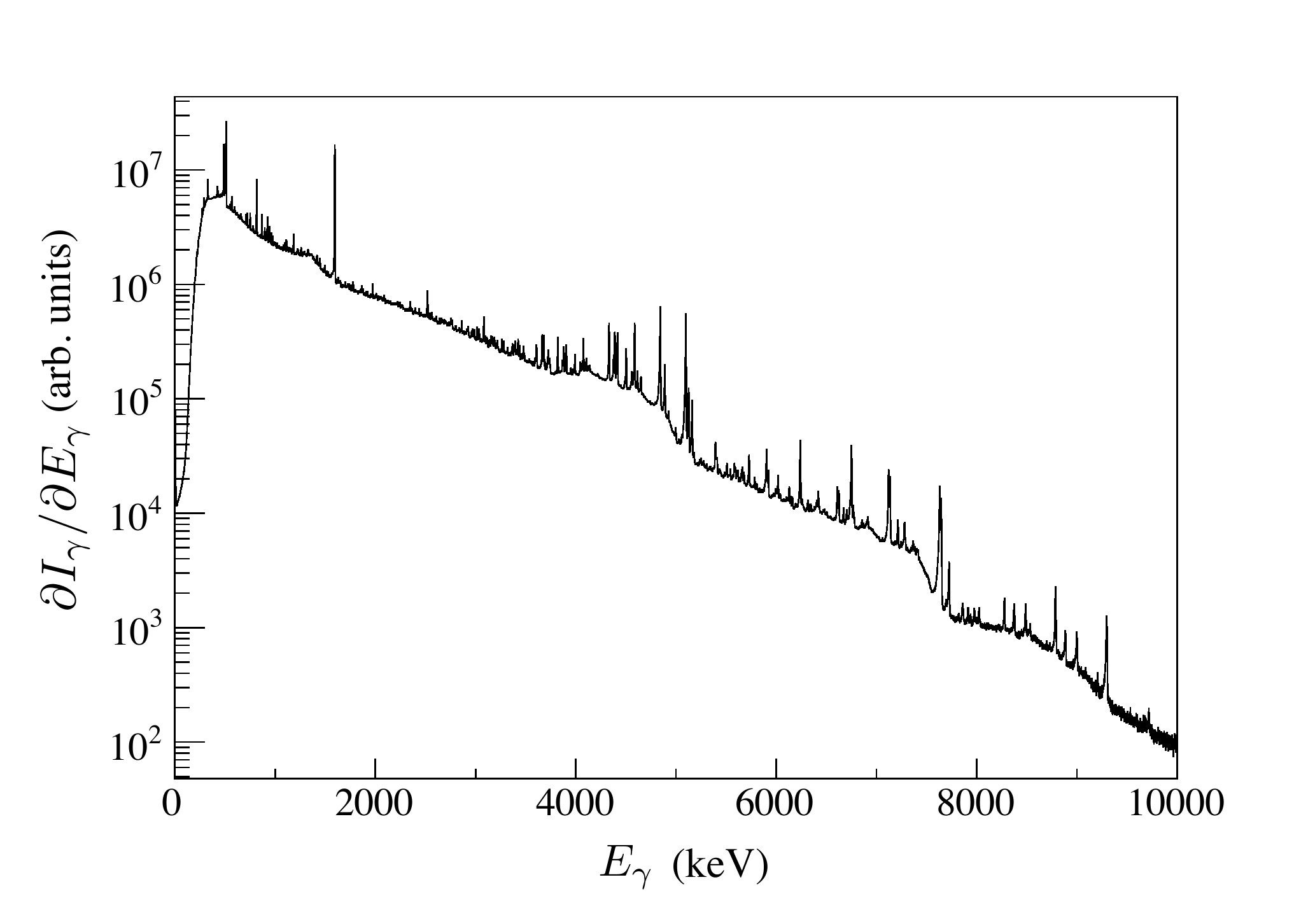}
  \caption{Pulse-height spectrum of $\gamma$ rays from the $(n,\gamma)$ reaction with a lanthanum target.}
  \label{gamma}
\end{figure}

The transverse asymmetry using inclusive $\gamma$-ray transitions and single $\gamma$-ray transitions were analyzed.
The neutron TOF spectrum and $\gamma$-ray spectrum are shown in Figs.\ref{tof} and \ref{gamma}, respectively. The total number of $\gamma$-ray events is denoted $I_{\gamma}$.

The number of protons injected into the spallation source was used to normalize the number of incident neutrons for each measurement.

The neutron transmittance of the ${\rm {^{3}He}}$ spin filter has time dependence due to the relaxation of ${\rm {^{3}He}}$ polarization. Therefore, the neutron transmittance obtained from Eq.\ref{transpol} was used to correct these effects. Since the neutron polarization also changed for the same reason, the correction was made using the neutron polarization determined by Eq.\ref{polpol}.

The full-absorption peak efficiency of the germanium detectors was normalized using full-absorption peak counts at $E_{\gamma}$ = 5269~keV in the $^{14}{\rm {N}}(n,\gamma)$ reaction with a melamine target. The reason for using $\gamma$ rays from the $^{14}{\rm {N}}(n,\gamma)$ reaction is that $\gamma$ rays emitted from this reaction do not have angular dependence.

The V1724 digitizer used in the data acquisition system for germanium detectors records only the time information of the input signal and loses the pulse height information when the interval between two signals becomes within 3.2~$\mu$s. It was confirmed that approximately 4$\%$ of the pulse height information was lost in the vicinity of the $p$-wave resonance. Since these events did not lose time information, they can be used in the analysis as a correction. When the interval between the two signals becomes within 0.4~$\mu$s, signals are processed as one signal. These events were considered to be approximately 0.4$\%$. The uncertainties due to these phenomena were negligible because they were sufficiently smaller than the statistical uncertainty.

The $\gamma$-ray yield for the $i$-th detector in the $p$-wave resonance region for incident up- and down- spin neutrons can be written as, considering the small cross section the vicinity of the $p$-wave resonance, respectively,

\begin{equation}
n^{\pm}_{n\gamma}(\theta _{i},\varphi_{i}) \propto \int_{E_{\gamma}}^{ }dE_{\gamma}\int_{t^{m}}^{ }dt^{m}\int_{\Omega_{i}}^{ } d\Omega \, \sigma^{\pm}_{n\gamma}(\theta,\varphi)\varepsilon_{i},
\end{equation}
where $\theta_{i}$ and $\phi_{i}$ are the $i$-th detector's mounting angle, $\Omega_{i}$ is the $i$-th detector's solid angle, $\varepsilon_{i}$ is the $i$-th detector's $\gamma$-ray detection efficiency evaluated from the $\gamma$-ray yield of the $^{14}{\rm {N}}(n,\gamma)$ reaction. In this analysis, the integral region in the $p$-wave is defined as $E_{p}-2\Gamma_{p} \leqslant  E_{n} \leqslant E_{p}+2\Gamma_{p}$ (See Table 1 for each definition and value.).
 
 \begin{table}[]
\caption{Resonance parameters of $^{139}{\rm {La}}$ used in the analysis. $r$ is the type of partial wave, $E_{r}$ is resonance energy, $J_{r}$ is the spin of compound state, $l_{r}$ is orbital angular momentum of the incident neutron, $\Gamma^{{\gamma}}_{r}$ is the partial $\gamma$ width, $g_{r} = \left ( 2J_{r}  +1  \right )/\left [2 \left ( 2I+1 \right ) \right ]$ is the statistical factor and $\Gamma^{n}_{r}$ is neutron width.}
\begin{tabular}{cccccc}
\hline
$r$ & $E_{r}$[eV]  & $J_{r}$ & $l_{r}$ & $\Gamma_{r}^{\gamma}$[meV] & $g_{r}\Gamma_{r}^{n}$[meV] \\ \hline
$s_{1}$ & $-48.63^{a}$ & $4^{a}$ & 0 & $62.2^{a}$ & $(571.8)^{a}$ \\
$p$ & $0.740\pm0.002^{b}$   & 4 & 1 & $40.41\pm0.76^{b}$ & $\left ( 5.6\pm0.5 \right )\times 10^{-5\: c}$ \\
$s_{2}$ & $72.30\pm0.05^{c}$   & 3 & 0 & $75.64\pm2.21^{c}$ & $11.76\pm0.53^{c}$ \\ \hline
\end{tabular}
 \begin{tablenotes}[para,flushleft]
   $^{a}$Taken from Ref.\cite{Mughabghab,JENDL}.\\
   $^{b}$Taken from Ref.\cite{Okudaira}.\\
$^{c}$Taken from Ref.\cite{nTOF}.
  \end{tablenotes}
\end{table}
 
The $\gamma$-ray yield in each detector was weighted using the detector position and A-type detectors were added to give the following expressions,

\begin{equation}
N^{\pm}={\sum_{i}}^{'}\frac{n^{\pm}_{n\gamma}(\theta _{i},\varphi_{i})}{\varepsilon_{i}\lambda_{i}} ,
\label{yeild_1}
\end{equation}

where the $i$-th detector's angular weight factor $\lambda_{i}$ is defined as $-\sin\theta_{i}\sin\varphi_{i}$ and ${\sum_{i}}^{'}$ represents the summation of lower A-type detectors.

The transverse asymmetry $A'^{\rm all}_{\textup{LR}}$ between $N^{\textup{+}}$ and $N^{\textup{-}}$ is defined as follows,
\begin{equation}
A'^{\rm all}_{\textup{LR}}= \frac{ N^{\textup{+}}-N^{\textup{-}} }{ N^{\textup{+}}+N^{\textup{-}} }.
\label{yeild_2}
\end{equation}
Here, when only $p$-wave resonance was focused, background component in the vicinity of the $p$-wave resonance, derived from the $s$-wave resonance and resonances of other nuclei were subtracted from Eq.\ref{yeild_1}. The transverse asymmetry $A'_{\textup{LR}}$ was then defined as in Eq.\ref{yeild_2}.

Neutron scattering in the target prior to absorption in the $p$-wave resonance alters the neutron momentum vector and hence impacts the asymmetry. Taking such scattering events into account, $A'_{\rm{LR}}$ is given by:
\begin{equation}
A'_{\rm{LR}}=\frac{\left (N^{+}_{s=0} + N_{s\geq 1}   \right ) -\left (N^{-}_{s=0} + N_{s\geq 1}  \right ) }{\left (N^{+}_{s=0} +N_{s\geq 1}  \right )+\left (N^{-}_{s=0} + N_{s\geq 1}  \right )},
\end{equation}
where $N_{s=0}^{\pm}$ and $N_{s\geq 1}$ represent the number of events that emit $\gamma$ rays without scattering in the target and the number of events that emit $\gamma$ rays after scattering in the target, respectively. Therefore, $A'_{\rm{LR}}$ and the true asymmetry $A_{\rm{LR}}$ can be related by:

\begin{equation}
A_{\rm{LR}}=A'_{\rm{LR}}\left ( 1+\frac{N_{s\geq 1}}{N_{r}} \right ),
\end{equation}
where $N_{r}$ is defined as $\frac{N^{+}_{s=0}+N^{-}_{s=0}}{2}$ ,which is equal to the component of the $p$-wave resonance. The ratio was estimated using a Monte Carlo simulation and estimated to be 0.0722 in the vicinity of the $p$-wave resonance.

\subsubsection{Inclusive $\gamma$-ray transition}
\begin{figure}[h]
  \centering
  \includegraphics[width=9cm]{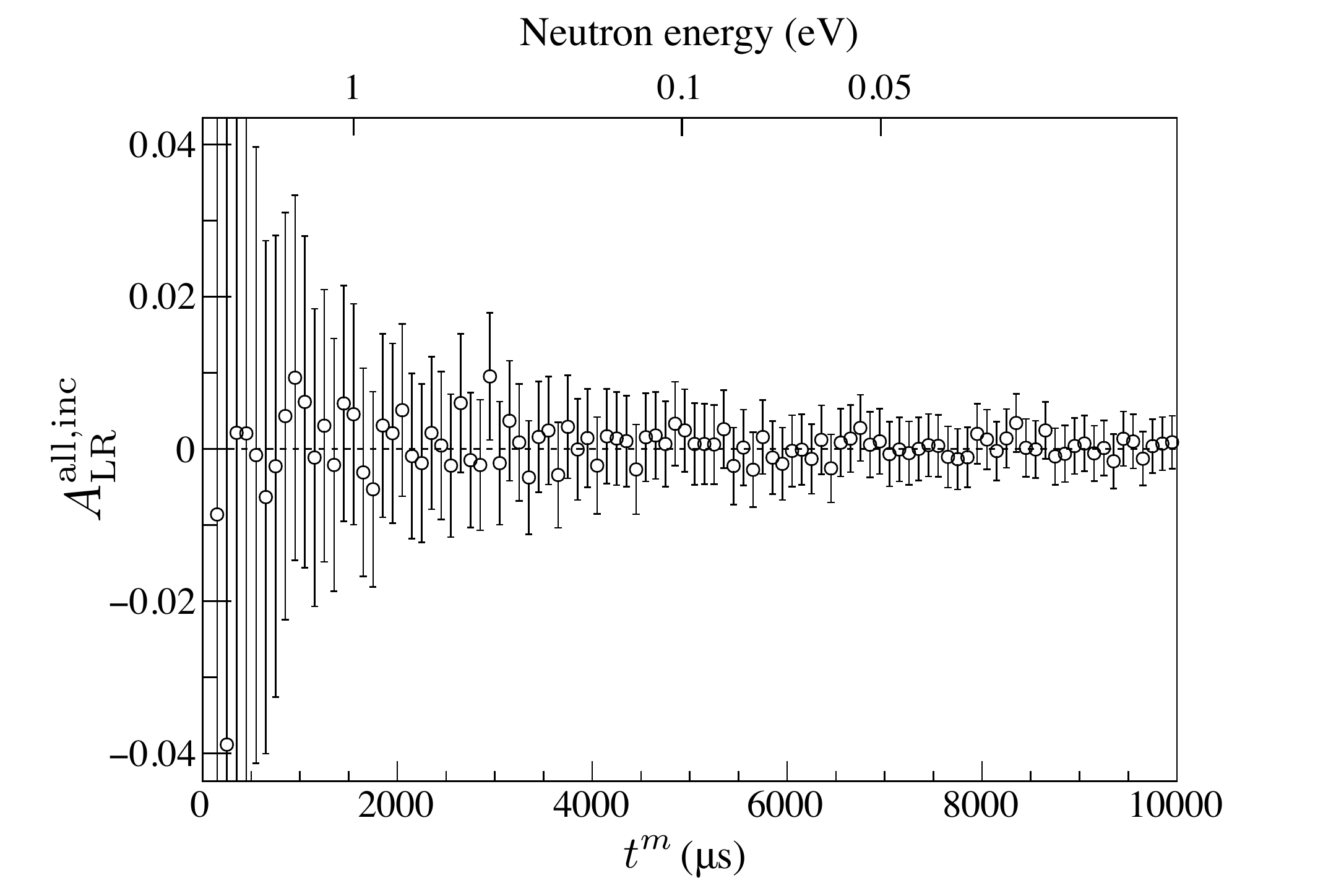}
  \caption{White points show the asymmetry for the all region when the threshold of the $\gamma$-ray was set to 2000~keV - 15000~keV.}
  \label{inc_asym}
\end{figure}

The asymmetry in the resonance region and continuous regions was evaluated by setting the threshold for $E_{\gamma}$ to 2000~keV - 15000~keV. The reason why the threshold is set to 2000~keV is to avoid the effect of delayed $\gamma$ rays with time-dependent yield. As shown in Fig.\ref{inc_asym}, it was confirmed that there was no asymmetry in the entire region.

\begin{figure}[h]
  \centering
  \includegraphics[width=9cm]{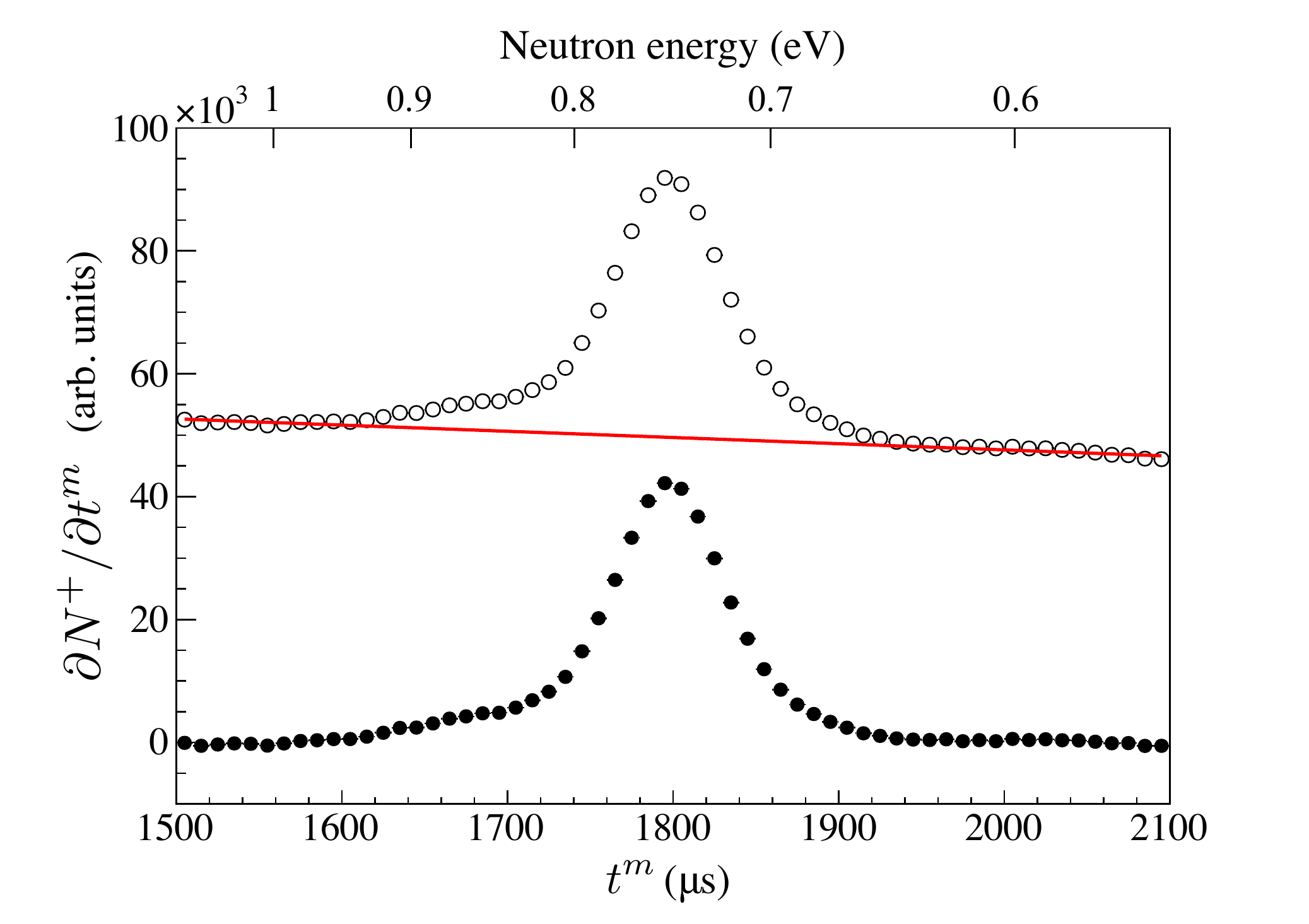}
  \caption{White points show the $\gamma$-ray yield in the vicinity of the $p$-wave resonance for up-polarized neutrons. Solid line shows fitting results to background. Black points indicate $p$-wave components after subtraction of background.}
 \label{bg_2000overup}
\end{figure}

Figure \ref{bg_2000overup} shows the $\gamma$-ray yield in the vicinity of the $p$-wave resonance. The background was subtracted by fitting using linear functions with $1400~\mu{\rm s} \leqslant t^{m} \leqslant 1600~\mu{\rm s}$, $1900~\mu{\rm s} \leqslant t^{m} \leqslant 2200~\mu{\rm s}$.

\begin{figure}[h]
  \centering
  \includegraphics[width=9cm]{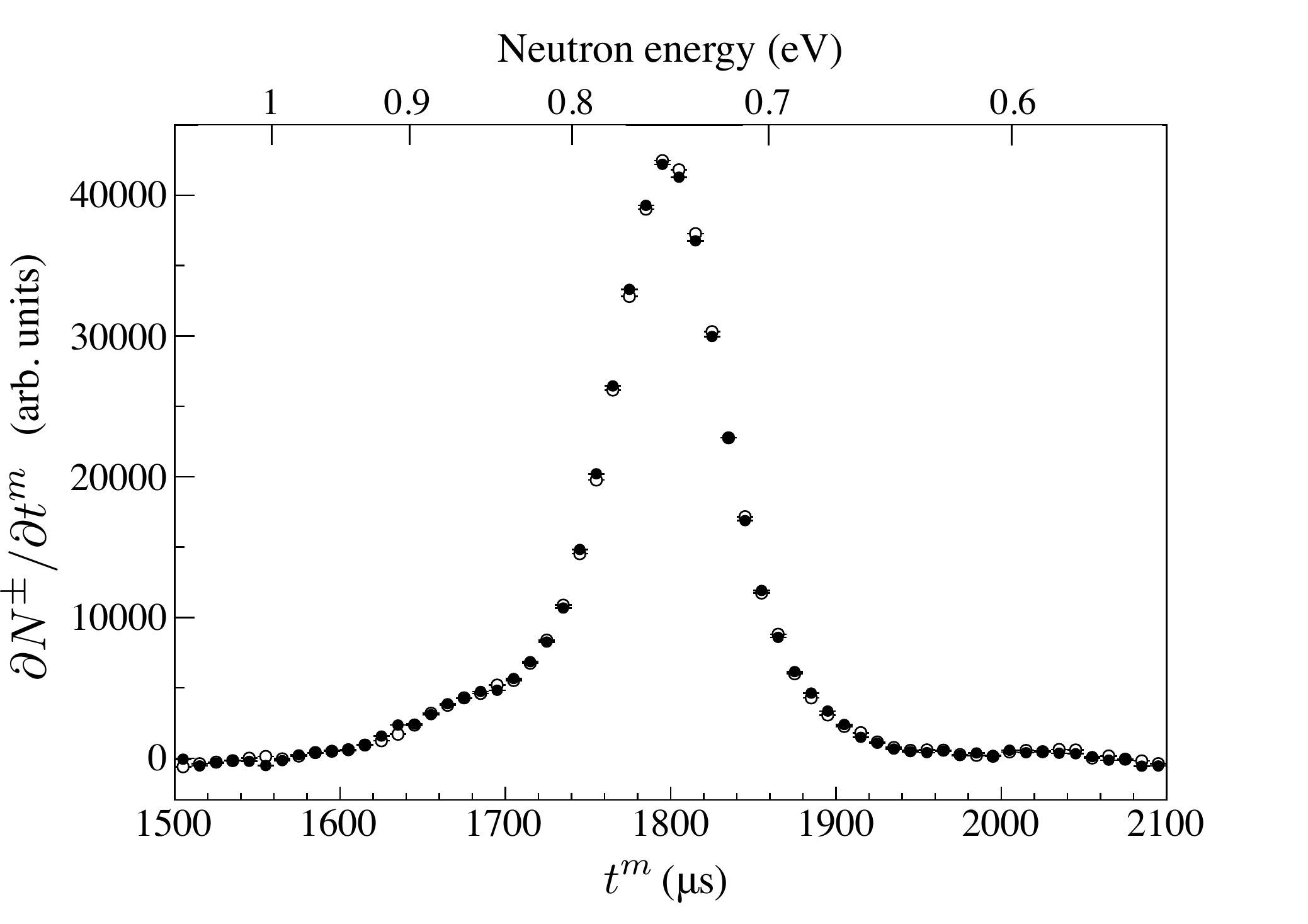}
  \caption{$\gamma$-ray yield in the vicinity of the $p$-wave resonance for each polarization direction of incident neutrons.
White and black points indicate the up- and down- polarization, respectively.}
 \label{2000over}
\end{figure}

Figure \ref{2000over} shows $\gamma$-ray yield in the vicinity of the $p$-wave resonance according to the spin state of incident neutrons after background subtraction. The value of the transverse asymmetry $A_{\textup{LR}}^{\rm {inc}}$ for inclusive $\gamma$-ray transitions was found to be $A_{\textup{LR}}^{\rm {inc}} = 0.0045 \pm 0.0080$.

\subsubsection{Single $\gamma$-ray transition}
The asymmetry for a single $\gamma$-ray transition was evaluated by applying the same procedure. 

\begin{figure}[h]
  \centering
  \includegraphics[width=9cm]{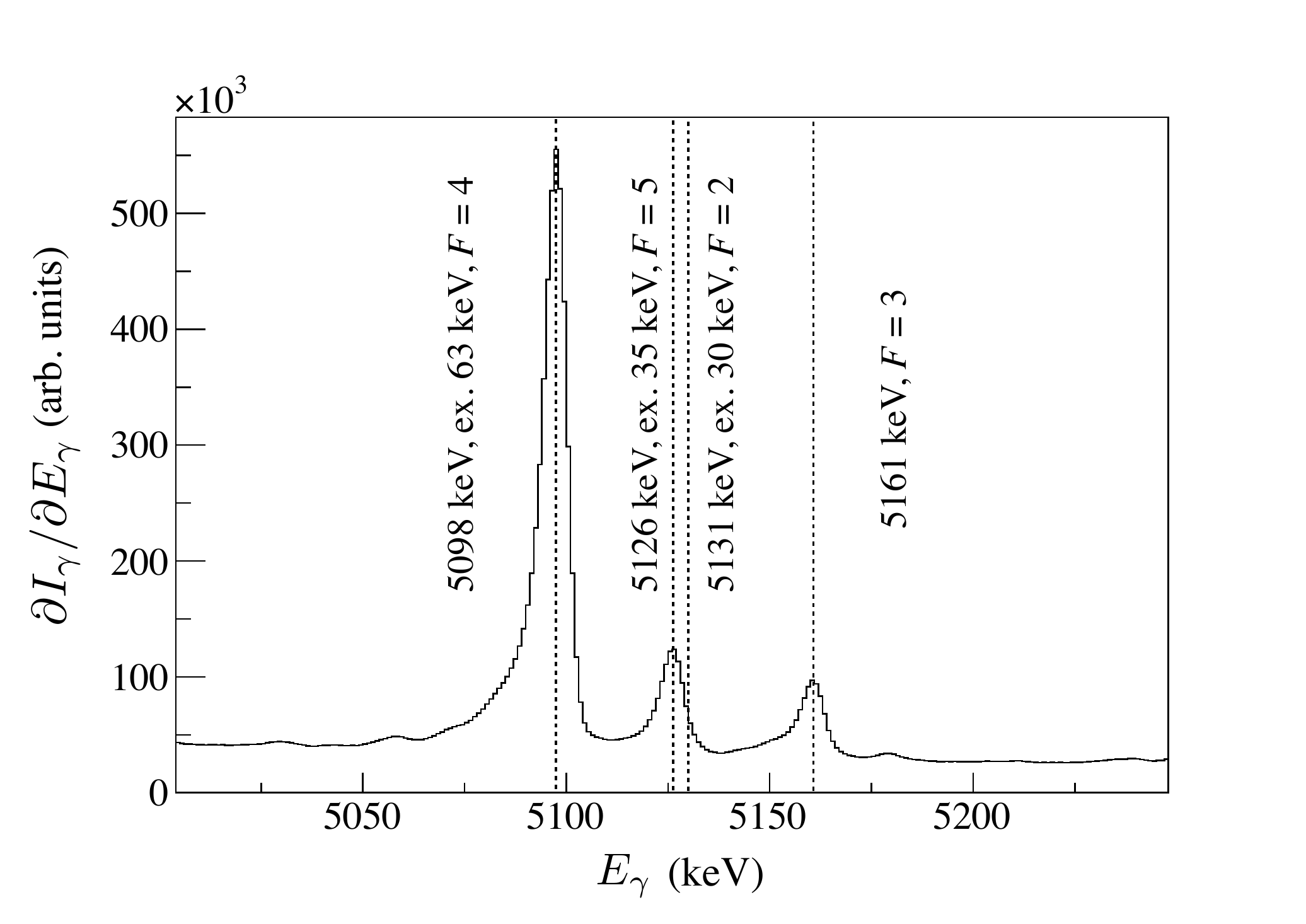}
  \caption{Pulse-height spectrum of $\gamma$-ray from $^{139}{\rm {La}}(n,\gamma)$ reaction, in particular, a detailed drawing around 5161~keV. Only the 5161~keV peak was used in this analysis.}
  \label{5161spec}
\end{figure}

In Fig.\ref{5161spec}, the peak at $E_{\gamma} = $ 5161~keV is the transition from the compound state of $^{139}{\rm {La}} + n$ to the ground state of $^{140}{\rm {La}}$ (spin of the final state: $F = 3$). This 5161~keV transition is the highest energy transition from $p$-wave resonance in the $^{139}{\rm {La}}(n,\gamma)$ reaction and free from the background induced from same reactions. The gate range was full width at quarter maximum (FWQM) of the full-absorption peak of the 5161~keV.

\begin{figure}[h]
  \centering
  \includegraphics[width=9.5cm]{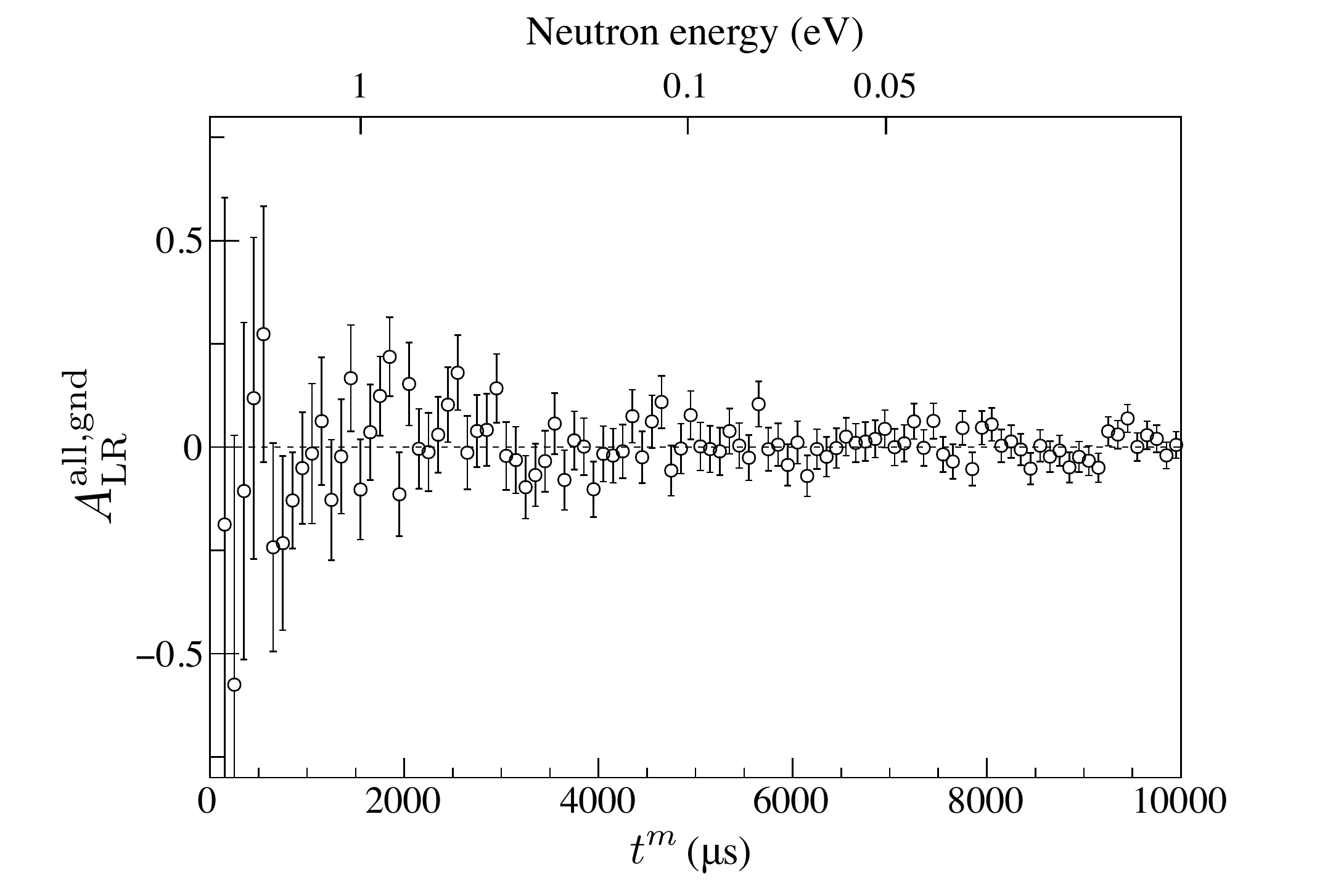}
  \caption{White points show the asymmetry for the entire region for the 5161~keV single $\gamma$-ray transition.}
  \label{5161_con_asym}
\end{figure}

\begin{figure}[h]
  \centering
  \includegraphics[width=9.5cm]{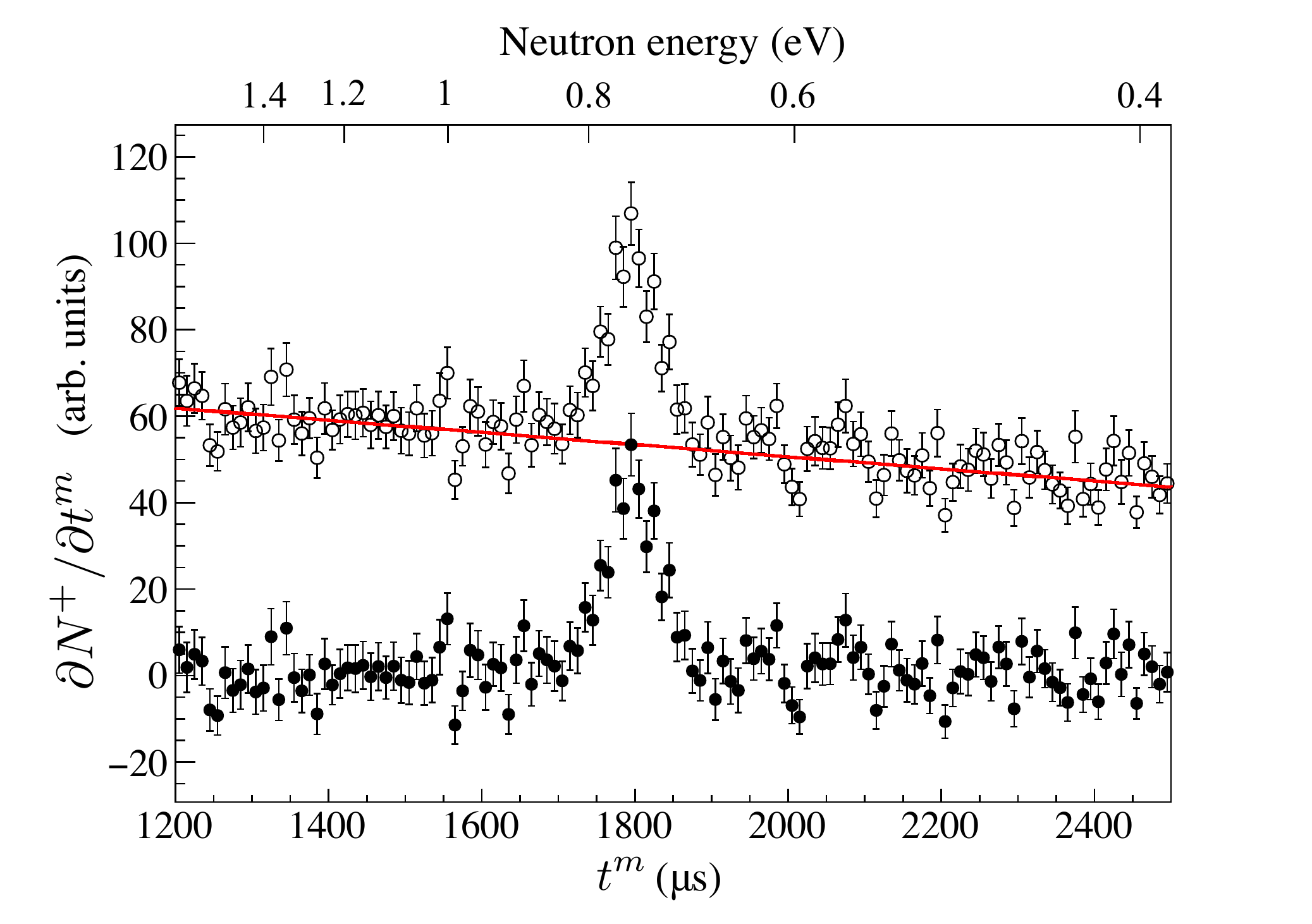}
  \caption{White points show the $\gamma$-ray yield in the vicinity of the $p$-wave resonance for up-polarized neutrons. Solid line shows fitting results to background. Black points indicate $p$-wave components after subtraction of background.}
  \label{bg_5161up}
\end{figure}

\begin{figure}[h]
  \centering
  \includegraphics[width=9.5cm]{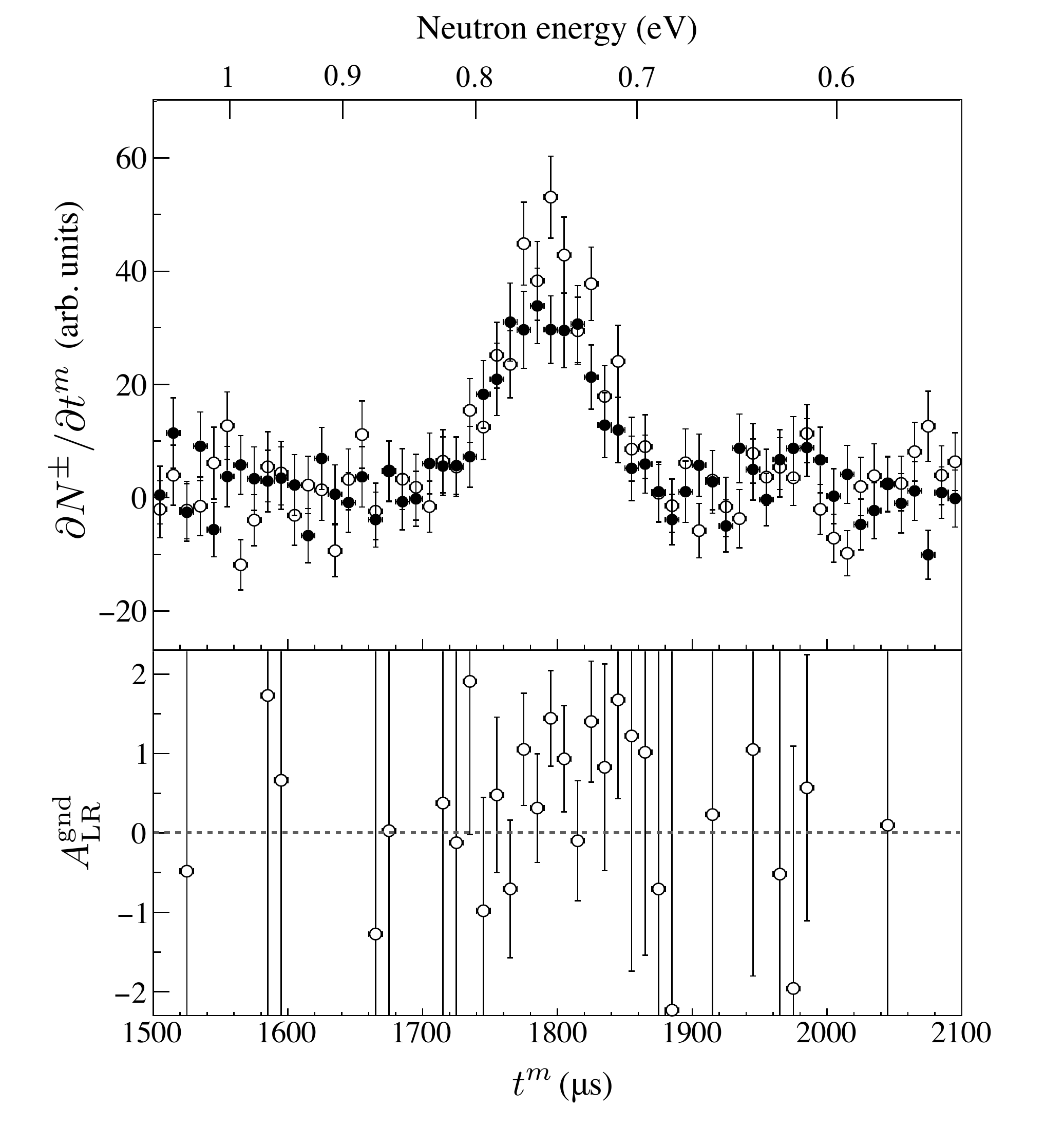}
  \caption{$\gamma$-ray yield in the vicinity of the $p$-wave resonance for each polarization direction of incident neutrons and the transverse asymmetry TOF spectrum.
 In the upper side figure, white and black points indicate the up- and down- polarization, respectively.}
 \label{5161}
\end{figure}

The neutron energy dependence of the asymmetry for the 5161~keV single $\gamma$-ray transition was evaluated. As shown in Fig.\ref{5161_con_asym}, it was confirmed that there was no asymmetry in the region excepted $p$-wave resonance. 
The asymmetry was evaluated focusing on the $p$-wave region. The $s$-wave component was evaluated by fitting linear functions with $1100~\mu{\rm s} \leqslant t^{m} \leqslant 1600~\mu{\rm s}$, $1900~\mu{\rm s} \leqslant t^{m} \leqslant 3300~\mu{\rm s}$ and subtracted as shown Fig.\ref{bg_5161up}.
Figure \ref{5161} shows $\gamma$-ray yield in the vicinity of the $p$-wave resonance according to the spin state of incident neutrons after background subtraction and the transverse asymmetry TOF spectrum. The value of the transverse asymmetry for 5161~keV single $\gamma$-ray transition was found to be $A_{\textup{LR}}^{\rm {gnd}} = 0.60 \pm 0.19$

\section{CONCLUSION}
This paper reported the first results of the transverse asymmetry of $\gamma$ rays in the $^{139}{\rm {La}}(\vec{n},\gamma)$ reaction using epithermal polarized neutrons. 
At J-PARC MLF BL04 ANNRI, it was demonstrated that the angular distribution of $\gamma$ rays in the $(\vec{n},\gamma)$ reaction using polarized neutrons could be measured.
The asymmetry of $\gamma$ rays was analyzed under two conditions. The value of the asymmetry for inclusive $\gamma$ rays was zero consistent, while there was the asymmetry only at the $p$-wave resonance for a single $\gamma$-ray transition and its value is $A_{\textup{LR}}^{\rm {gnd}} = 0.60 \pm 0.19$. \\
In the near future, we will comprehensively analyze the asymmetry and the results of previous studies using the framework of the $sp$-mixing model.

\section{ACKNOWLEDGEMENTS}
The authors would like to thank the staff of ANNRI for the maintenance of the germanium detectors, and MLF and J-PARC for operating the accelerators and the neutron-production target. The neutron experiments at the Materials and Life Science Experimental Facility of J-PARC were performed under the user program (Proposals No. 2018B0148, No. 2019A0185). This work was supported by the Neutron Science Division of KEK as an S-type research project with program number 2018S12. This work was partially supported by MEXT KAKENHI Grant No. JP19GS0210 and JSPS KAKENHI Grant No. JP17H02889.












\end{document}